\documentclass[prl,noshowkeys,twocolumn,superscriptaddress]{revtex4}
\usepackage{graphicx}
\usepackage{amsmath}

\def\gap{\;\rlap{\lower 2.5pt
 \hbox{$\sim$}}\raise 1.5pt\hbox{$>$}\;}
\def\lap{\;\rlap{\lower 2.5pt
   \hbox{$\sim$}}\raise 1.5pt\hbox{$<$}\;}
\def\gsim{\;\rlap{\lower 2.5pt
 \hbox{$\sim$}}\raise 1.5pt\hbox{$>$}\;}
\def\lsim{\;\rlap{\lower 2.5pt
   \hbox{$\sim$}}\raise 1.5pt\hbox{$<$}\;}
\def\msun{{\rm\,M_\odot}}

\def\spose#1{\hbox to 0pt{#1\hss}}
\def\lta{\mathrel{\spose{\lower 3pt\hbox{$\mathchar''218$}}
     \raise 2.0pt\hbox{$\mathchar''13C$}}}
\def\gta{\mathrel{\spose{\lower 3pt\hbox{$\mathchar''218$}}
     \raise 2.0pt\hbox{$\mathchar''13E$}}}
\newcommand{\be}{\begin{equation}}
\newcommand{\ee}{\end{equation}}

\newcommand{\ls}{\mathrel{\raise1.16pt\hbox{$<$}\kern-7.0pt 
\lower3.06pt\hbox{{$\scriptstyle \sim$}}}}         
\newcommand{\gs}{\mathrel{\raise1.16pt\hbox{$>$}\kern-7.0pt 
\lower3.06pt\hbox{{$\scriptstyle \sim$}}}}         

\long\def\comment#1{}

\def\mh{M_{\bullet}}
\def\msun{M_{\odot}}
\def\fun#1#2{\lower3.6pt\vbox{\baselineskip0pt\lineskip.9pt
  \ialign{$\mathsurround=0pt#1\hfil##\hfil$\crcr#2\crcr\sim\crcr}}}
\def\lap{\mathrel{\mathpalette\fun <}}
\def\gap{\mathrel{\mathpalette\fun >}}
\newcommand{\ba}{\begin{eqnarray}}
\newcommand{\ea}{\end{eqnarray}}

\begin{document}
\bibliographystyle{apsrev.bst}
\title{Evolution of the Dark Matter Distribution at the Galactic Center}
\author{David Merritt}
\affiliation{Department of Physics, Rochester Institute of Technology,
Rochester, NY 14623}

\begin{abstract}
Annihilation radiation from neutralino dark matter at the Galactic
center (GC) would be greatly enhanced if the dark matter
were strongly clustered around the supermassive black hole (SBH).
The existence of a dark-matter ``spike''
is made plausible by the observed, steeply-rising {\it stellar} density
near the GC SBH.
Here the time-dependent equations describing gravitational
interaction of the dark matter with the stars are solved.
Scattering of dark matter particles by stars would
substantially lower the dark matter density near the GC SBH
over $10$ Gyr, 
due both to kinetic heating, and to capture of 
dark matter particles by the SBH.
This evolution implies a decrease by several orders of magnitude in
the observable flux of annihilation products compared with 
models that associate a steep dark matter spike with the SBH.
\end{abstract}

\maketitle

Neutralinos in supersymmetry are likely candidates for the non-baryonic
dark matter \cite{PP:82,Goldberg:83}.
If neutralinos make up a large fraction of the dark matter in the
galactic halo, pair annihilations will produce an excess of photons
which may be observed in gamma ray detectors \cite{Bergstrom:99}.
The galactic center (GC) is a promising target for such
searches since the dark matter density is predicted to
rise as $\rho\sim r^{-1}$ at the centers of dark matter 
halos \cite{Navarro:03}.
In addition, the GC contains a supermassive black hole
(SBH) with mass $\mh\approx 10^{6.5}\msun$ \cite{Schoedel:03}.
``Adiabatic growth'' models in which the SBH remains 
stationary as it grows predict 
the formation of a steep power-law density profile
around the SBH, a ``spike,'' and an increase by many
orders of magnitude in the amplitude of the neutralino 
annihilation signal \cite{Gondolo:99}.

The bulges of galaxies like the Milky Way are believed to have
formed via mergers of pre-existing stellar systems.
If the latter contained SBHs, a merger would result in the
formation of a binary SBH \cite{BBR:80}.
The density of stars and dark matter around a binary SBH
drops rapidly as the binary ejects matter via the
gravitational slingshot \cite{Quinlan:96}.
Even a binary with mass ratio as extreme as $1:10$ would
efficiently destroy a dark matter spike on parsec scales
\cite{Merritt:02a}.

Evidence for the scouring effect of binary SBHs is seen
at the centers of the brightest galaxies 
\cite{Milos:02,Ravindranath:02}, 
where the {\it stellar}
density profiles are nearly flat and sometimes even
exhibit a central minimum \cite{Lauer:02}.
However in fainter elliptical galaxies and in the bulges of spiral
galaxies like the Milky Way, 
steeply-rising stellar densities are observed:
$\rho_\star\sim r^{-\gamma}, 1.5\lap\gamma\lap 2.5$.
In these galaxies, the most recent mergers may have taken place
before the era at which SBHs formed, allowing the stellar
density near the SBH to remain high.
Since stars and dark matter respond similarly to the
presence of a SBH, galaxies with steeply-rising
stellar densities are the most plausible sites for steeply-rising
dark matter densities and hence for the detection of annihilation
radiation  \cite{Silk:02}.

Stars near the SBH would also interact with the dark matter
via gravitational scattering \cite{Ilyin:03,Gnedin:03}.
Here, the time-dependent equations describing the scattering of
dark matter particles off of stars in the presence of a SBH are
solved.
Scattering decreases the density of a dark matter spike by
kinetic heating, and by driving particles into the SBH.
The result, after $10$ Gyr, is the virtual dissolution
of the dark matter spike in a galaxy like the Milky Way.
This result suggests that enhancements in the dark matter density
around the GC SBH would be modest whether or not the Milky Way
bulge has experienced the scouring effects of a binary SBH.

Let $r_h$ be the radius of gravitational influence of the
SBH, with $M_*(r<r_h)=2\mh$.
For the GC SBH, $r_h\approx 1.67$ pc \cite{Genzel:03}.
After growth of the SBH (assumed to remain fixed with respect
to the bulge),
the dark matter density is approximately
\begin{subequations}
\begin{eqnarray}
\rho(r)  = \rho(r_b)&\times & (r/r_b)^{-\gamma_{sp}},\ r\lap r_b \\
	&\times& (r/r_b)^{-\gamma_c},\ r\gap r_b
\end{eqnarray}
\end{subequations}
where $\gamma_{sp} = 2 + 1/(4-\gamma_c)$,
$\rho\propto r^{-\gamma_c}$ is the dark matter density
before growth of the SBH, and $r_b\approx 0.2 r_h$ \cite{Merritt:04}.

An estimate of the local heating rate of dark matter particles 
due to gravitational encounters with stars is
the change per unit time of $\epsilon=\frac{1}{2}mv_{rms}^2$,
the mean kinetic energy of the dark matter particles.
Assuming Maxwellian velocity distributions for the stars and
dark matter,
\begin{equation}
\frac{d\epsilon}{dt}
=\frac{8\left(6\pi\right)^{1/2} G^2\rho_\star m\ln\Lambda}{\left(v_{rms}^2 + 
v_{\star,rms}^2\right)^{3/2}} (\epsilon_\star - \epsilon)
\label{eq_equipart}
\end{equation}
where $\epsilon_\star={1\over 2}m_\star v_{\star,rms}^2$,
$\rho_\star$ is the stellar mass density and $\ln\Lambda$
is the Coulomb logarithm \cite{Spitzer:87}.
Taking the limit $m\ll m_\star$ and assuming $v_{rms}\approx v_{\star,rms}$,
appropriate shortly after the dark matter spike forms,
the local heating time becomes
\begin{subequations}
\begin{eqnarray}
T_{\rm local} &\equiv& \left| {1\over\epsilon} {d\epsilon\over dt}\right|^{-1} = {0.0814 v_{rms}^3\over G^2m_\star\rho_\star\ln\Lambda}\\
&\approx& 1.8\times 10^9 {\rm yr} \left({r\over 1{\rm pc}}\right)^{-0.1};
\end{eqnarray}
\end{subequations}
the latter expression uses the observed stellar mass density near the 
GC SBH, 
$\rho_\star \approx 3.2\times 10^5\msun {\rm pc}^{-3}(r/1{\rm pc})^{-\gamma}$,
$\gamma = 1.4\pm 0.1$ \cite{Genzel:03}, and 
$v_{\star,rms}\approx 1.12 (G\mh/r)^{1/2}$ with $\mh=3\times 10^6\msun$.
The Coulomb logarithm was set to $\ln\Lambda=\ln(0.4N)$ with
$N\approx 6\times 10^6$ the number of stars within $r_h$
\cite{Spitzer:71}.
The time to heat the dark matter is nearly independent of radius 
and shorter by a factor $\sim 5$ than the age 
($\sim 10$ Gyr; \cite{Zoccali:03}) of the stellar bulge.

The change with time of the dark matter density can be computed
from the Fokker-Planck equation describing the evolution of 
$f({\bf r},{\bf v}, t)$, the mass density
of dark matter particles in phase space, due to gravitational
interactions with stars \cite{Merritt:83}.
We assume that $f$ is isotropic in velocity space,
$f=f(E,t)$, with $E\equiv -v^2/2 +\phi(r)$
the binding energy per unit mass of a dark matter particle, 
and $\phi(r)=-\Phi(r)$, with $\Phi(r)$ the gravitational potential due
to the SBH and the stars.  
The kinetic equation describing the evolution of $f(E,t)$ due to scattering
off of stars with masses $\gg m$ is
\begin{subequations}
\begin{eqnarray}
& & 4\pi^2p(E){\partial f(E,t)\over\partial t} = -{\partial F_E\over\partial E} - F_{lc}(E,t) ,\\
& & F_E(E,t) = -D_{EE}(E){\partial f\over\partial E},\\
& & D_{EE}(E) = 64\pi^4G^2\ln\Lambda\times  \nonumber \\ 
& & \left[q(E) \int_{-\infty}^E dE'h_\star(E') + 
\int_E^\infty dE' q(E')h_\star(E')\right].
\end{eqnarray}
\label{eq:diffeq}
\end{subequations}
Here $p(E) = 4\sqrt{2}\int_0^{r_{max}(E)} dr r^2 \sqrt{\phi(r)-E}
=-\partial q/\partial E$
is the phase space volume accessible per unit of energy,
with $\phi(r_{max}) = E$.
The mass density of dark matter particles in $E$-space
is $N(E)dE=4\pi^2 p(E)f(E)dE$.
The dark matter heating rate is determined by 
$h_\star(E)=\int \ m_\star f_\star(E,m_\star) dm_\star$ 
with $f_\star(E,m_\star) dm_\star$ the mass density 
of stars in phase space in the interval $m_\star$ to
$m_\star + dm_\star$.
If the distribution $n(m_\star)dm_\star$ of stellar masses is assumed
independent of energy (i.e. distance from the black hole), then
$h_\star(E)=\tilde m_\star f_\star(E)$, with $f_\star(E)$ the
total mass density of stars in phase space and
\begin{equation}
\tilde m_\star = {\int n(m_\star) m_\star^2 dm_\star\over 
\int n(m_\star) m_\star dm_\star}.
\label{eq:meff}
\end{equation}
$F_{lc}(E)$ is the flux of stars that are scattered 
from low angular momentum orbits into the SBH
and is discussed in more detail below.
Eqs. (\ref{eq:diffeq}) assume that small-angle scatterings 
dominate the evolution
of $f$ and that the gravitational potential changes on a time scale
long compared with $T_{\rm local}$; 
the latter assumption is valid as long as the
gravitational acceleration is not produced dominantly by the dark
matter particles themselves.

Neglecting $F_{lc}$,
the total energy ${\cal E}=\int_{E_1}^{E_2} N(E)EdE$
of dark matter particles in the energy range $E_1<E<E_2$
changes with time according to
\begin{subequations}
\begin{eqnarray}
{d{\cal E}\over dt} &=& -\int_{E_1}^{E_2} dE\ E\ {\partial F_E\over\partial E} \\
&=& -\left[EF_E\right]_{E_1}^{E_2} - \left[fD_{EE}\right]_{E_1}^{E_2} - \\
& & \int_{E_1}^{E_2} dE N(E) Q(E),\\
Q(E)&=& 16\pi^2 G^2\tilde m_\star\ln\Lambda \int_0^E dE' f_\star(E').
\end{eqnarray}
\label{eq:qofe}
\end{subequations}
The third term in eq. (\ref{eq:qofe}c), 
which is always negative, represents heating of the dark matter.
We accordingly define the (non-local) time scale for heating of the dark 
matter particles to be
\begin{equation}
T_{\rm heat}^{-1} \equiv 
{\int_{E_1}^{E_2} dE N(E) Q(E)\over \int_{E_1}^{E_2} dE EN(E)}.
\end{equation}
This expression may be used to estimate the dissolution time of
a dark matter spike.
Assume that both stars and dark matter particles 
initially have power-law density profiles near the SBH: 
$\rho(r,t=0) \propto r^{-\gamma_{sp}}$,
$\rho_\star(r) \propto r^{-\gamma}$, $r\lap r_h$,
and that $\phi(r)=G\mh/r$.
The isotropic distribution function corresponding to an $r^{-\gamma}$
density profile in an $r^{-1}$ potential is $f(E)\propto E^{\gamma-3/2}$.
Setting $E_1=\phi(r_h)=G\mh/r_h$ and $E_2\rightarrow\infty$,
$T_{\rm heat}$ becomes
\begin{subequations}
\begin{eqnarray}
& & T_{\rm heat} = A(\gamma,\gamma_{sp}){\mh\over \tilde m_\star} 
\left({G\mh\over r_h^3}\right)^{-1/2} {1\over\ln\Lambda},\\
& & A(\gamma,\gamma_{sp}) = \nonumber \\
& & {1\over 2} \sqrt{\pi\over 3} {(\gamma-1/2)(7/2-\gamma-\gamma_{sp})\over (3-\gamma)(2-\gamma_{sp})} {\Gamma(\gamma+1/2)\over\Gamma(\gamma+1)}.
\end{eqnarray}
\label{eq:heat}
\end{subequations}
When $\gamma=3/2$, equal within the uncertainties with the slope of 
the stellar cusp around the Milky Way SBH \cite{Genzel:03},
the coefficient $A(\gamma,\gamma_{sp})$ 
in eq. (\ref{eq:heat}) is independent of $\gamma_{sp}$
and
\begin{subequations}
\begin{eqnarray}
& & T_{\rm heat} = {4\sqrt{3}\over 27} {\mh\over \tilde m_\star} 
\left({G\mh\over r_h^3}\right)^{-1/2} {1\over\ln\Lambda} \\
& & = 1.25\times 10^9 {\rm yr}\times \nonumber \\
& & \left({\mh\over 3\times 10^6\msun}\right)^{1/2} 
\left({r_h\over 2\ {\rm pc}}\right)^{3/2} 
\left({\tilde m_\star\over\msun}\right)^{-1} \left({\ln\Lambda\over 15}\right)^{-1}.
\end{eqnarray}
\label{eq:theat}
\end{subequations}
The effective stellar mass $\tilde m_\star$ that appears in equation
(9a) depends via equation (\ref{eq:meff})
on the mass function $n(m_\star)$
of stars in the GC stellar cusp.
While $n(m_\star)$ is not strongly constrained by observations,
either at the high or low mass ends,
it is sometimes assumed (e.g. \cite{Genzel:03}) to be 
a power law with Salpeter \cite{Salpeter:55} index, 
$n(m_\star) \propto m^{-(1+\alpha)}$,
$\alpha\approx 1.35$.
Setting $m_{\star,min}=0.08\msun$ and $m_{\star,max}=(5)10(20)\msun$
then yields $\tilde m_\star\approx (0.8)1.2(1.8)\msun$.

We take $T_{\rm heat}$ as defined in eq. (\ref{eq:theat}) 
as our unit of time in what follows,
with $\tau\equiv t/T_{\rm heat}$.
The age of the majority of the stars near the GC 
is $\gap 10$ Gyr \cite{Zoccali:03},
although some much younger ($t\lap 10^7$ yr) stars are 
present in the cusp at distances $\lap 0.1$ pc from
the SBH \cite{Genzel:00,Ghez:03b}.
Setting $t=10$ Gyr, $r_h=1.67$ pc 
and $1\le \tilde m_\star/\msun\le 2$ gives
$10\lap\tau\lap 20$.
If the young stellar population is continually replenished, 
$m_\star$ and $\tau$ could be larger, implying a higher
mean rate of dark matter heating.

\begin{figure}
\includegraphics[height=12cm]{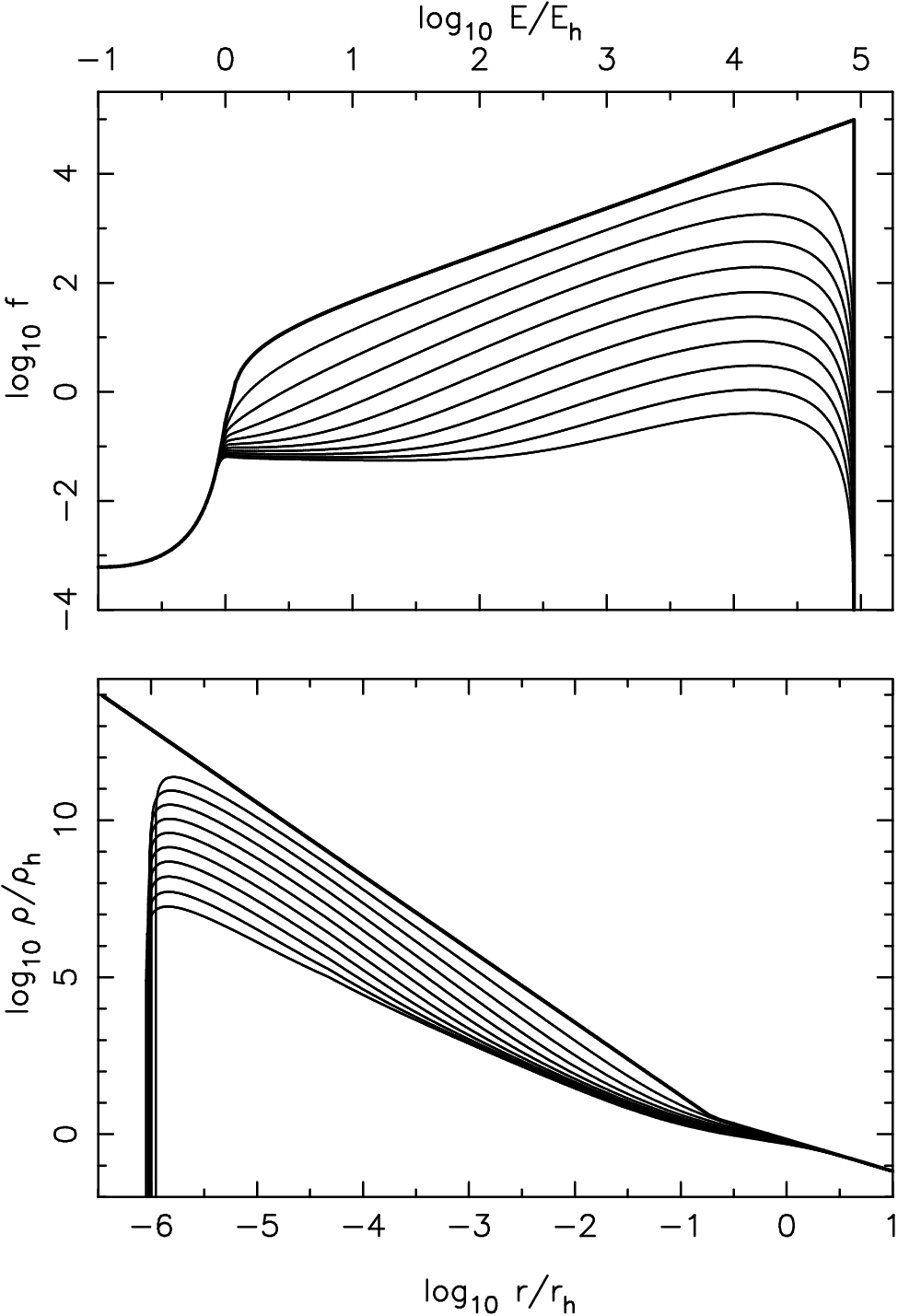}
\caption{Evolution of the dark matter phase space density $f(E)$ and
mass density $\rho(r)$ due to gravitational interactions with stars
around the GC SBH.
The initial dark matter density is given by eq. 1 with $\gamma_{sp}=7/3$
and $\gamma_c=1$.
Times shown are $\tau=0$ (heavy curves) and $\tau = 2, 4, ..., 20$
where $\tau$ is the time in units of $T_{\rm heat}$ (eq. \ref{eq:theat});
$10\le\tau\le 20$ corresponds roughly to the age of the galactic bulge.
}
\label{fig1}
\end{figure}

Diffusion in energy will cause a modest loss of stars to the SBH,
$\dot M = -F_E(E_2)$, $E_2\approx c^2$.
A much greater capture rate is implied by scattering
of dark matter particles on low angular momentum (eccentric)
orbits into the SBH \cite{FR:76}.
The loss rate is given approximately by \cite{LS:77}
\begin{subequations}
\begin{eqnarray}
F_{lc}(E) &\approx& 4\pi^2P(E)J_c^2(E)\overline{\mu}(E) R{\partial f\over\partial R}\\
&\approx& S(E) f(E),\\
S(E) &=& 4\pi^2 P(E)J_c^2(E)\overline{\mu}(E) 
\left[\ln R_0(E)^{-1}\right]^{-1}.
\end{eqnarray}
\label{eq:CK}
\end{subequations}
Here $R\equiv J^2/J_c^2(E)$ is a scaled angular momentum variable
with $J_c(E)$ the angular momentum of a circular orbit of energy $E$;
$P$ is the period of a radial orbit;
$\overline{\mu}$ is the orbit-averaged angular momentum diffusion
coefficient $\langle(\Delta R)^2\rangle/2R$;
and $R_0$ is the value of the angular momentum variable
at which $f$ drops to zero due to the competing effects of capture
and diffusion.
The final line of eq. (\ref{eq:CK}) assumes $f\sim\ln[R/R_0(E)]$ 
near the loss cone \cite{LS:77,CK:78}.
Cohn \& Kulsrud \cite{CK:78} give expressions for $R_0$ as a function
of $E$ based on solutions to the $R-$dependent Fokker-Planck equation;
we adopt their expressions here.
The angular momentum diffusion coefficients used by these authors,
for modelling systems containing a single stellar mass, may
be shown to remain unchanged when the scattered objects
(here dark matter) have masses much less that those of the scatterers (stars).
The rate of loss of stars predicted by eq. (\ref{eq:CK}) depends only weakly
on the radius of the capturing sphere, which we set
to $2G\mh/c^2$.
The diffusion coefficient $\overline{\mu}$ is of order 
$T^{-1}_{\rm heat}$, hence 
$F_{lc}(E)\approx N(E)/T_{\rm heat}(E)\ln R_0^{-1}$.

The detailed evolution of the dark matter density around the
GC SBH was computed by integrating eq. (\ref{eq:diffeq}a) forward
in time.
The stellar density was modelled via Dehnen's \cite{Dehnen:93}
density law,
$\rho_\star(r) \propto (r/r_0)^{-\gamma}(1+r/r_0)^{\gamma-4}$
with $\gamma=1.4$ and $r_0$
chosen to match the observed stellar density at $r\lap r_h$ \cite{Genzel:03}.
Fig. 1 shows the evolution of the dark matter density assuming
$\gamma_{sp}=7/3$ and $\gamma_c=1$, the values corresponding
to a spike that developed in response to adiabatic growth of
a SBH in a dark matter halo with $\gamma_c=1.0$.
Fig. 1 shows that scattering of dark matter particles by stars causes the
dark matter density to drop and the spike to flatten,
within a radius $\sim r_h$ where the heating
time is shorter than the age of the bulge.
Fig. 2 shows the dark matter density at $\tau=10$
for a range of initial spike profiles,
$1.0\le\gamma_{sp}\le 2.75$.
The mean density within
$r_h$ drops by several orders of magnitude when
$\gamma_{sp}\approx 2$.
When $\gamma_{sp}=\gamma_c=1$, i.e. no spike, the net effect of the
heating is to increase the dark matter density slightly.

\begin{figure}
\includegraphics[height=8cm]{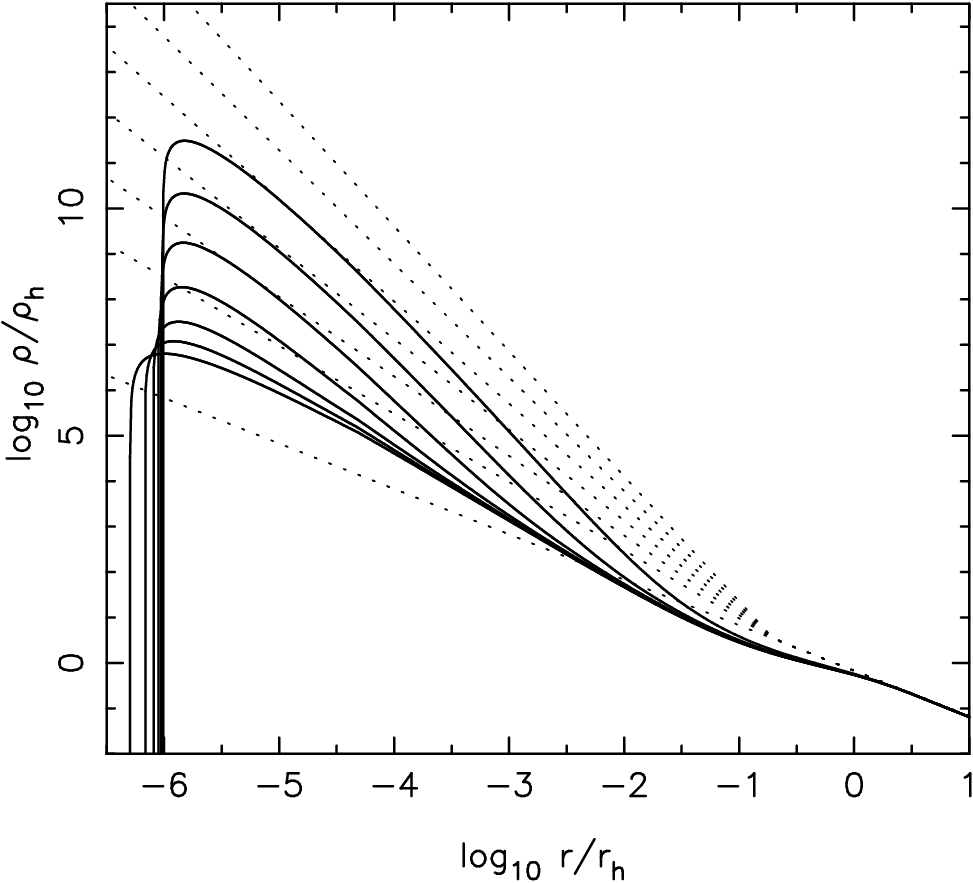}
\caption{Dark matter density at $\tau=0$
(dotted lines) and $\tau=10$ (solid lines) 
given an initial density that satisfies
eq. 1, with  $\gamma_c=1$ and
$\gamma_{sp}=(1.00,1.50,1.75,2.00,2.25,2.50,2.75)$.
}
\label{fig2}
\end{figure}

In the absence of scattering into the SBH,
eqs. (\ref{eq:diffeq}) have the time-independent
solution $f(E)=$ constant, $\rho\sim r^{-3/2}$ \cite{Gnedin:03}.
As Figs. 1 and 2  show, there is a tendency to evolve
toward this characteristic profile, although a number of factors
keep it from being precisely reached, 
including the finite evolution time; 
the presence of the loss term $F_{lc}$; 
and the fact that $f={\rm constant}$ can only
hold true over a finite range of energies given the 
boundary conditions on $f$.
Nevertheless, at late times ($\tau\gap 20$),
the solutions found here are generally well described
by $\rho\sim r^{-3/2}$ at radii $10^{-5}\lap r/r_h\lap 10^{-2}$.

The flux of dark matter annihilation photons
along a direction that makes 
an angle $\psi$ with respect to the GC is
proportional to the line-of-sight integral
$\int_{\psi} \rho^2 dl$.
Following earlier authors \cite{Bergstrom:99},
we define the dimensionless form factor
$J(\psi)\equiv K\int_{\psi}\rho^2(l) dl$,
$K^{-1}=(8.5\ \mathrm{kpc})(0.3\ \mathrm{GeV/cm}^3)$.
Given a photon detector with angular acceptance
$\Delta\Omega$ directed toward the GC,
the signal is proportional to
\begin{equation}
\langle J\rangle \equiv {1\over \Delta\Omega} \int_{\Delta\Omega}J(\psi)\ 
d\Omega.
\label{eq:defj}
\end{equation}
Figure 3 shows the evolution of $\langle J\rangle$
for $\Delta\Omega = 10^{-5}(10^{-3})$ sr;
the first value is the approximate solid angle
of the detectors in GLAST \cite{GLAST} and in atmospheric Cerenkov
telescopes like VERITAS \cite{VERITAS}, while the larger angle corresponds
approximately to EGRET \cite{EGRET}.
The dark matter density was normalized to a fiducial value of
$\rho=100\msun {\rm pc}^{-3}$
at $r=r_h$; $J$ and $\langle J\rangle$ scale as $\rho^2(r_h)$.
We note that $\rho(r_h)$ is uncertain and could be much 
lower \cite{Ullio:01,Merritt:02a,Evans:03}.
Figure 3 shows that the very large initial values of $\langle J\rangle$
are rapidly diminished as the spike is dissolved;
by $\tau=10$, $\langle J\rangle$ has dropped below $\sim (10^4,10^3)$,
$\Delta\Omega=(10^{-5},10^{-3})$ for all $\gamma_{sp}\lap 2.5$.
These values are similar to what would be predicted for the central
regions of a dark matter halo in the absence of a SBH \cite{Bergstrom:99}.

\begin{figure}
\includegraphics[height=11cm]{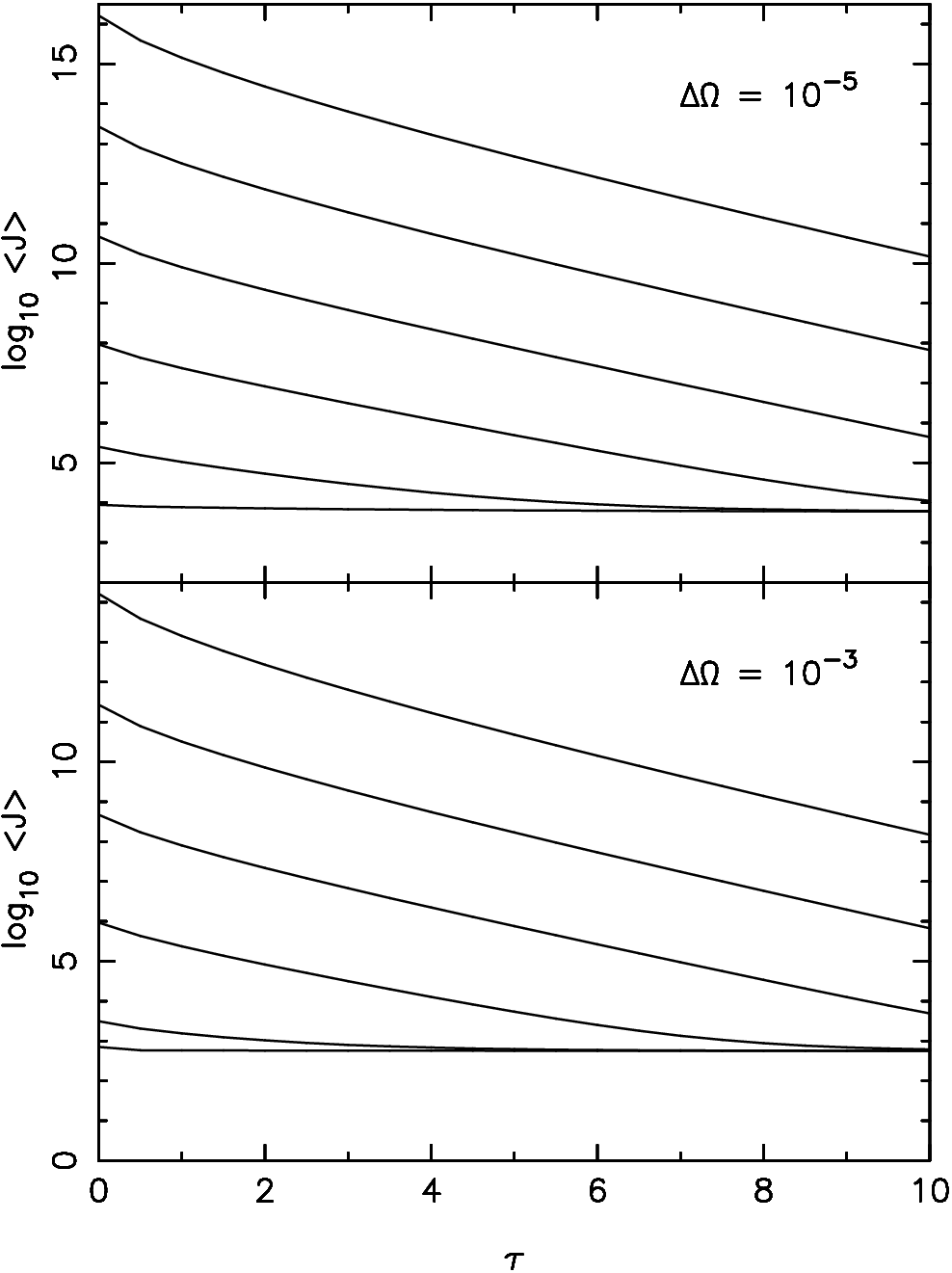}
\caption{Evolution of $\langle J\rangle$, 
a dimensionless measure of the cuspiness of the dark
matter spike (eq. \ref{eq:defj}), for two values of the
solid angle $\Delta\Omega$ of a detector centered on
the SBH.
Curves are shown for $\gamma_{sp}=1.50$ (lower), $1.75,2.00,2.25,2.50$,
and $2.75$ (upper).
The dark matter density was normalized to an initial value of
$100\msun {\rm pc}^{-3}$ at $r=r_h$;
$\langle J\rangle$ scales as $\rho^2(r_h)$.
}
\label{fig3}
\end{figure}

The dark mass captured by the SBH after $10$ Gyr
is less than $10^4\msun$ ($\rho(r_h)=100\msun {\rm pc}^{-3}$)
for all the integrations presented here.
Various schemes have been discussed for increasing the captured
mass in stars or dark matter to much greater values, perhaps of order $\mh$.
These include making the dark matter collisional \cite{Ostriker:00};
assuming instantaneous replenishment of the loss cone 
\cite{MacMillan:02,Zhao:02};
or allowing the stellar potential to be non-axisymmetric \cite{Merritt:03b}.
The first two mechanisms are ad hoc; the third, if it applies to the
GC, might allow the persistence of a dark matter spike in the face
of scattering and capture by increasing the mass in dark matter
particles that can interact with the SBH.

I thank Milos Milosavljevic for useful discussions.
This work was supported by NSF grant AST 02-0631 
and by NASA grant NAG5-9046.

\end{document}